\documentclass[aps,twocolumn,pra,superscriptaddress,amsmath,showpacs,tightenlines]{revtex4-1}

\usepackage{amsmath}
\usepackage{amssymb}
\usepackage{graphicx}
\usepackage{xcolor}
\usepackage{hyperref}

\begin{document}

\title{Detection of emitter-resonator coupling strength in quantum Rabi model via an auxiliary resonator}
\author{Xinyun Cui}
\affiliation{Center for Quantum Sciences and School of Physics, Northeast Normal University,
Changchun 130024, China}
\affiliation{Center for Advanced Optoelectronic Functional Materials Research, and Key
Laboratory for UV-Emitting Materials and Technology of Ministry of
Education,Northeast Normal University, Changchun 130024, China}
\author{Zhihai Wang}
\email{wangzh761@nenu.edu.cn}
\affiliation{Center for Quantum Sciences and School of Physics, Northeast Normal University,
Changchun 130024, China}
\affiliation{Center for Advanced Optoelectronic Functional Materials Research, and Key
Laboratory for UV-Emitting Materials and Technology of Ministry of
Education,Northeast Normal University, Changchun 130024, China}
\author{Yong Li}
\email{liyong@csrc.ac.cn}
\affiliation{Beijing Computational Science Research Center, Beijing 100193, China}
\affiliation{Synergetic Innovation Center for Quantum Effects and Applications, Hunan Normal University, Changsha 410081, China}

\begin{abstract}
In this paper, we propose a theoretical scheme {to detect} the emitter-resonator coupling strength in the ultra-strong coupling regime in the quantum Rabi model via introducing an auxiliary resonator. We demonstrate the total system as a two-mode Rabi model and obtain the ground state by the transformed rotating wave approximation, which is {shown to be} superior to the usually applied rotating wave approximation. Here, {the coupling strength} is detected by monitoring the average excitation number in the auxiliary resonator and the sensitivity of the detection scheme is discussed {analytically}.
\end{abstract}

\maketitle

\section{Introduction}
The quantum Rabi model~\cite{Rabi} describes the interaction between a two-level emitter and a {single} bosonic  mode in quantum optics. In the early {days, the rotating wave approximation (RWA) {was} usually applied, and the Rabi model is approximated as a Jaynes-Cummings model~\cite{JC}}. However, the recent experimental progress has made it possible to access the ultra-strong and deep-strong coupling regime~\cite{ultra1,ultra2,ultra3,ultra4,ultra5,ultra6,ultra7,ultra10}, {where the coupling strength is comparable or larger than the frequency of the bosonic mode or/and the emitter and} the effect of counter rotating wave {terms} plays a crucial role. Therefore, the studies of quantum Rabi model beyond the RWA are becoming a hot topic in recent years.

Generally speaking, the recent investigations {on} the quantum Rabi model mainly cover the following aspects: (I)  {Seeking for the exact solutions of the Rabi and anisotropic (mixed) Rabi models by use of the Bargmann algebra and Bogoliubov operators~\cite{exact1,exact2,exact3,exact4}}. (II) {Accessing} the analytical solutions by resorting to various approximations. For example, the generalized rotating wave approximation~\cite{GRWA1,GRWA2}, the transformed rotating wave approximation (TRWA)~\cite{TRWA1,TRWA2, TRWA3,TRWA4}, the generalized squeezing rotating-wave approximation~\cite{squeeze1,squeeze2}, as well as the generalized variational approach~\cite{var1,var2,var3}. (III) Studying the quantum phase transition in the Rabi-type model, in the situation that the ratio of frequency of the two-level emitter to that of the bosonic mode tends to be infinity~\cite{phase1,phase2,phase3,phase4,phase5,phase6,phase7}. (IV) Simulating the ultra-strong coupling~\cite{phase7,simu1,simu2,simu3} and {investigating its potential applications~\cite{holo1,ultra10,zeno1,zeno2,zeno3}}. However, we find that there are {only a few discussions} about how to detect the emitter-field coupling strength sensitively~\cite{JL,Solano}.

{It is well known that}, a direct detection will undoubtedly disturb the detected quantum system. Therefore, we {here} introduce an auxiliary resonator, which {couples} to the resonator {containing} a two-level emitter, to perform the detection of the emitter-resonator coupling strength in the quantum Rabi model. Thus, the total system under consideration is demonstrated as a two-mode Rabi model~\cite{ultra5,twomode}. We generalize the TRWA {approach~\cite{TRWA1} to study our two-mode Rabi model}, and obtain the ground state approximately. {Unlike the case with RWA, where the ground state has zero excitation for the bosonic mode, the ground state in our model with counter rotating wave terms yields non-zero bosonic excitations. This paves} the way to perform the detection of the emitter-resonator coupling strength by monitoring the average excitation number in the auxiliary resonator. We find that a strong photonic hopping strength is beneficial for enhancing the sensitivity of our detecting strategy. Compared to the dynamical detection with the assistance of a auxiliary two-level system~\cite{JL,Solano}, our scheme is based on the ground state of the quantum system {and it would be} robust to the unavoidable system-environment interaction.

The rest of the paper is organized as follows. In Sec.~\ref{model}, we introduce our two-mode Rabi model and discuss the available experimental setup. In Sec.~\ref{method}, we generalize the TRWA approach to obtain the ground state approximately, and make a comparison with that under the RWA. Based on the ground state, we discuss the sensitivity of coupling strength detection by monitoring the average excitation number in the auxiliary resonator in Sec.~\ref{detect}. At last, we give a conclusion in Sec.~\ref{conclusion}.

\section{Model and Hamiltonian}
\label{model}

\begin{figure}[tbp]
\centering
\includegraphics[width=1\columnwidth]{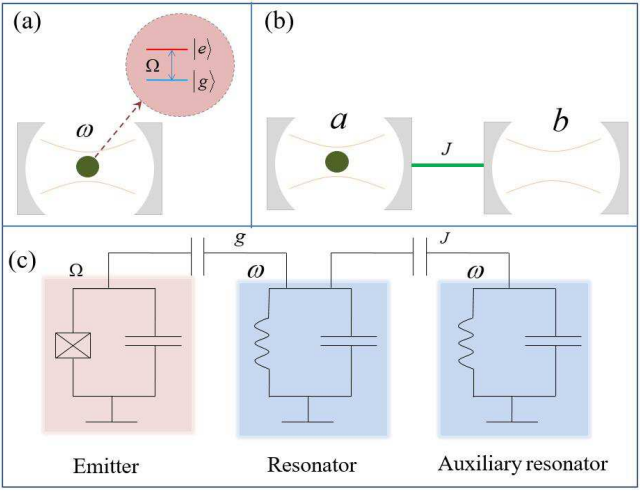}
\caption{(Color online) Schematic configuration of the detection proposal. (a) The two-level emitter interacting with a single-mode resonator. (b) The coupling to the auxiliary resonator which is used  for performing the detection of the emitter-resonator coupling. (c) The effective circuit diagram of the device.}
\label{scheme}
\end{figure}

As schematically shown in Fig.~\ref{scheme}(a), the quantum Rabi model describes the interaction between a two-level emitter and a single bosonic mode, which {can be} supplied by a resonator. The corresponding Hamiltonian reads ($\hbar=1$)
\begin{equation}
H_R=\omega a^{\dagger}a+\frac{\Omega}{2}\sigma_{x}+\frac{g}{2}(a^{\dagger}+a)\sigma_{z},
\label{Rabi}
\end{equation}
{where $\Omega$ is the energy-level splitting of the two-level emitter, $\omega$ is the frequency of the bosonic mode, and $g$ is the coupling strength.} $\sigma_{x}$ and $\sigma_{z}$ are the Pauli matrices to describe the emitter, with $\sigma_x|e\rangle=|e\rangle,\sigma_x|g\rangle=-|g\rangle$ {$\sigma_z|g\rangle=|e\rangle, |\sigma_z|e\rangle=|g\rangle$} and $a^\dagger$ $(a)$ is the creation (annihilation) operator of the bosonic mode.

In the nearly resonant and strong coupling regime, the RWA is usually applied and the Hamiltonian in Eq.~(\ref{Rabi}) takes the form of the Jaynes-Cummings model~\cite{JC}. However, in the ultra-strong or deep-strong coupling or large detuning regime, the RWA breaks down and the resonator mode will acquire non-zero excitations {even} in the ground state. To perform the detection of the coupling strength $g$, we introduce an auxiliary resonator as shown in Fig.~\ref{scheme}(b), which couples to the original resonator mode, and the total Hamiltonian {can be written as a two-model Rabi model~\cite{ultra5,twomode}}
\begin{eqnarray}
H&=&H_R+H_d,\label{HH}\\
H_{d}&=&\omega b^{\dagger}b+J(a^{\dagger}b+b^{\dagger}a),
\end{eqnarray}
where $H_d$ is the Hamiltonian for the detection scheme. Here, $b$ is the annihilation operator for the auxiliary resonator which supports a single bosonic mode being resonant with the resonator $a$. $J$ is the coupling strength between the two resonators.

Physically speaking, the two-mode Rabi model under consideration can be realized in circuit QED system as shown in Fig.~\ref{scheme}(c). Here, the two-level emitter is supplied by the transmon qubit and the LC circuit serves as the bosonic mode with frequency of $6$ -- $8$\,GHz~\cite{simu2}. The coupling between {the} two resonators has been realized in Ref.~\cite{catch}, where it is used to catch and release the microwave photons instead of performing detection here. In an alternative way, the bosonic mode can also be realized by the superconductive transmission line~\cite{ultra8,ultra9}, and the coupling strength between the qubit and the bosonic mode has been achieved in the ultra-strong and deep strong coupling regimes~\cite{ultra6,ultra7}.

\section{Approximate ground state}
\label{method}

To tackle the counter-rotating wave terms in the Hamiltonian, we {here} resort to the TRWA approach~\cite{TRWA1}. To this end, we firstly define a pair of super bosonic modes $A$ and $B$ as
\begin{eqnarray}
A=\frac{1}{\sqrt{2}}(a+b),\ \ B=\frac{1}{\sqrt{2}}(a-b),
\end{eqnarray}
which are symmetric and anti-symmetry superposition of the two local resonator modes. {In terms of $A$ and $B$}, the Hamiltonian of the whole system becomes
\begin{eqnarray}
 H&=&(\omega+J)A^{\dagger}A+(\omega-J)B^{\dagger}B+\frac{\Omega}{2}\sigma_{x}
 \nonumber \\&&+\frac{g}{2\sqrt{2}}(A+A^{\dagger}+B+B^{\dagger})\sigma_{z}.
 \label{Rabi2}
\end{eqnarray}
The Hamiltonian~(\ref{Rabi2}) describes that a two-level emitter interacts with two non-degenerate bosonic super modes simultaneously {[the frequency of $A$ ($B$) is $\omega+J$ ($\omega-J$)]}. To obtain the ground state of the system, we here generalize the TRWA approach which is originally proposed to deal with the traditional Rabi model with only one bosonic mode~\cite{TRWA1}. Following the spirit of the TRWA, we begin with the unitary transformation performing on the Hamiltonian by $H'=UHU^\dagger$, with
\begin{eqnarray}
U=\exp[{\xi_{A}(A^{\dagger}-A)\sigma_{z}+\xi_{B}(B^{\dagger}-B) \sigma_{z}}],
\end{eqnarray}
where the parameters $\xi_{A}$ and $\xi_{B}$ are real and will be determined later. After a cumbersome but direct calculation, the Hamiltonian $H'$ is obtained as $H'=H_0+H_1+H_2$, where
\begin{subequations}
\begin{eqnarray}
H_{0}&=&\frac{1}{2}\eta\Omega\sigma_{x}+(\omega+J)A^{\dagger}A+(\omega-J)B^{\dagger}B\nonumber\\
&&+(\omega+J)\xi_{A}^{2}+(\omega-J)\xi_{B}^{2}-\frac{g\xi_{A}}{\sqrt{2}}-\frac{g\xi_{B}}{\sqrt{2}},\\
H_{1}&=&[\frac{g}{2\sqrt{2}}-(\omega+J)\xi_{A}](A^{\dagger}+A)\sigma_{z}\nonumber\\
&&+[\frac{g}{2\sqrt{2}}-(\omega-J)\xi_{B}](B^{\dagger}+B)\sigma_{z}\nonumber\\
&&+\frac{i}{2}\eta\Omega\sigma_{y}[2\xi_{A}(A^{\dagger}-A)+2\xi_{B}(B^{\dagger}-B)],\\
H_{2}&=&\frac{1}{2}\Omega\sigma_{x}\{\cosh[2\xi_{A}(A^{\dagger}-A)+2\xi_{B}(B^{\dagger}-B)]-\eta\}\nonumber\\
&&+\frac{i}{2}\Omega\sigma_{y}\{\sinh[2\xi_{A}(A^{\dagger}-A)+2\xi_{B}(B^{\dagger}-B)]\nonumber\\
&&-\eta[2\xi_{A}(A^{\dagger}-A)+2\xi_{B}(B^{\dagger}-B)]\},
\end{eqnarray}
\end{subequations}
and $\eta$ is the vacuum average
\begin{eqnarray}
\eta&:=&\langle0_A,0_B|\cosh[2\xi_{A}(A^{\dagger}-A)+2\xi_{B}(B^{\dagger}-B)]|0_A,0_B\rangle\nonumber\\
&=&\exp[-2(\xi_{A}^{2}+\xi_{B}^{2})].
\end{eqnarray}

It is obvious that $H_0$ is exactly solvable, $H_1$ describes the effective linear interaction between the emitter and the bosonic modes, which is composed of both the rotating wave and counter rotating wave terms. The counter rotating wave terms in $H_1$ can be eliminated when $\xi_A$ and $\xi_B$ satisfy
\begin{eqnarray}
\xi_{A}=\frac{\sqrt{2}g}{4[(\omega+J)-\eta\Omega]},\, \xi_{B}=\frac{\sqrt{2}g}{4[(\omega-J)-\eta\Omega]}.
\label{parameter}
\end{eqnarray}
{and then the Hamiltonian $H_1$ can be reexpressed as
\begin{eqnarray}
H_1&=&\frac{\sqrt 2g\eta\Omega}{2(\eta\Omega-\omega-J)}(A^\dagger \sigma_-+A\sigma_+)\nonumber\\
&&+\frac{\sqrt 2g\eta\Omega}{2(\eta\Omega-\omega+J)}(B^\dagger \sigma_-+B\sigma_+),
\end{eqnarray}
where $\sigma_{\pm}=(\sigma_z\pm i\sigma_y)/2$.}

{Next, to obtain the approximate ground state, we will approximate the total Hamiltonian as  the TRWA Hamiltonian $H'\approx H_0+H_1$, by neglecting $H_2$ for the following two reasons: (i) It is obvious that $\langle 0_A,0_B,g|H_2|0_A,0_B,g\rangle=0$, where $|0_A,0_B,g\rangle:=|0\rangle_A\otimes|0_B\rangle\otimes|g\rangle$ is the ground state of $H_0$. (ii) The Hamiltonian $H_2$ only includes two- and multi-photon transitions, whose contributions to the physical quantity can be neglected~\cite{TRWA1}. In what follows, we will compare the approximate result based on the TRWA Hamiltonian $H'$ with the numerical results based on the exact Hamiltonian $H$ in Eq.~(\ref{HH}), to check the validity of our approach}.

{It is obvious that $H'$ has the similar form with the Jaynes-Cummings model, therefore, we can readily obtain the approximate ground state energy as}
 \begin{eqnarray}
E_g\approx-\frac{1}{2}\eta\Omega+(\omega+J)\xi_{A}^2-\frac{g\xi_{A}}
{\sqrt{2}}+(\omega-J)\xi_{B}^2-\frac{g\xi_{B}}{\sqrt{2}},\nonumber \\
\label{EG}
\end{eqnarray}
and the corresponding ground state
\begin{eqnarray}
|G_{T}\rangle&\approx&\exp{[-\xi_{A}(A^{\dagger}-A)\sigma_{z}
-\xi_{B}(B^{\dagger}-B)\sigma_{z}]}|0_A,0_B,g\rangle\nonumber\\
&=&\frac{1}{\sqrt{2}}\{\exp{[-\xi_{A}(A^{\dagger}-A)-\xi_{B}(B^{\dagger}-B)]}
|0_A,0_B,\uparrow\rangle\nonumber \\
&&-\exp{[\xi_{A}(A^{\dagger}-A)+\xi_{B}(B^{\dagger}-B)]}|0_A,0_B,\downarrow\rangle\},\nonumber\\
\label{ground}
\end{eqnarray}
where $\left|\uparrow\right\rangle$ and $\left|\downarrow\right\rangle$ are the eigen states of $\sigma_z$ with $\sigma_z\left|\uparrow\right\rangle=\left|\uparrow\right\rangle,
\sigma_z\left|\downarrow\right\rangle=-\left|\downarrow\right\rangle$.

\begin{figure}[tbp]
\centering
\includegraphics[width=1\columnwidth]{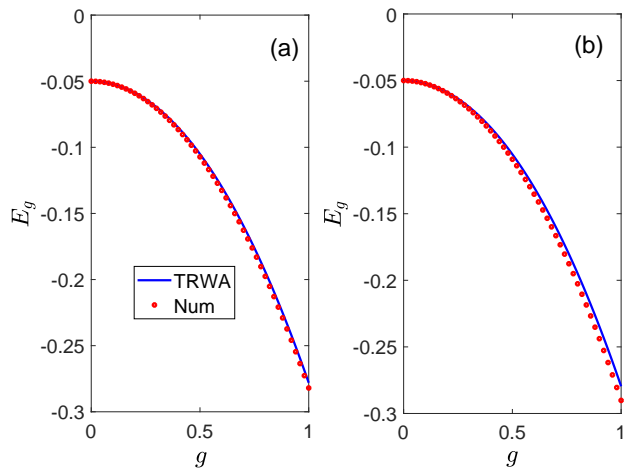}
\caption{(Color online) The ground state energy as a function of the coupling strength $g$. The solid blue line represents the approximate analytical results and the red empty circle is the numerical results. The parameters are set as $\Omega=0.1$, $\omega=1$, and (a) $J=0.05$, (b) $J=0.2$.}
\label{energy}
\end{figure}

To check the validity of our approximations, we compare the {approximate} ground state energy in Eq.~(\ref{EG}) with the exact results by directly diagonalizing the Hamiltonian of two-mode Rabi model in Eq.~(\ref{HH}). As shown in Fig.~\ref{energy}, in the parameter regime of $\Omega\ll\omega$, that is, the energy splitting of the emitter is much smaller than {the resonant frequency} of the bosonic mode, our approximate results agree well with the numerical results. As for the usual RWA, the ground state energy is readily given by $E_g^{\rm RWA}=-\Omega/2>E_g$, {independent of $g$ and $J$}, so that our approach has made a significant improvement beyond the RWA. {The reason of the improvement is that we have taken the displacement induced by the counter rotating wave terms into consideration via the unitary transformation $U$}. As shown in the wave function [Eq.~(\ref{ground})], the photonic counterpart of the ground state with counter rotating wave terms yields a coherent state with nonzero excitations, and this displacement with respect to the vacuum state under the RWA lowers the ground state energy, especially in the regime {of large $g$}.

\begin{figure}[tbp]
\centering
\includegraphics[width=1\columnwidth]{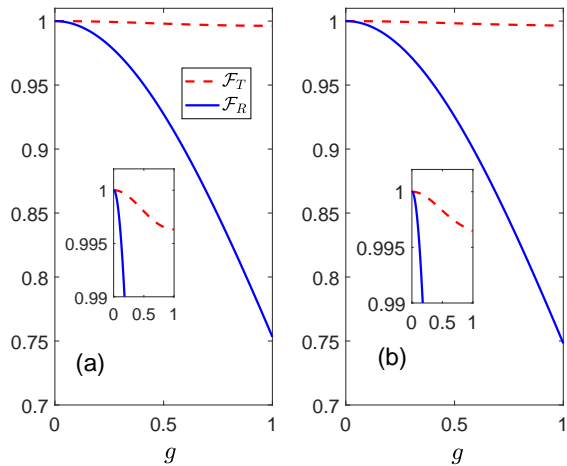}
\caption{(Color online) The fidelity of the ground state as a function of the coupling strength $g$. The red dashed lines represent our approximate analytical results based on TRWA and the solid blue lines are the results from RWA. The parameters are set as $\Omega=0.1$, $\omega=1$, and (a) $J=0.05$, (b) $J=0.2$.}
\label{fidelity}
\end{figure}

Furthermore, {we continue to check the validity of the approximate ground wave function by investigating its fidelity}. Here, the fidelity is defined as {$\mathcal{F}_T:=|\langle G|G_T\rangle|^2$}, where $|G\rangle$ is the ground state obtained by numerically diagonlizing the Hamiltonian in Eq.~(\ref{HH}), and $|G_T\rangle$ is the approximate result in Eq.~(\ref{ground}). In Fig.~\ref{fidelity}, we plot the fidelity as a function of the coupling strength $g$ for different $J$ ($J=0.05$ and $J=0.2$). It shows that the fidelity {can achieve} as high as $99.5\%$ in a broad parameter regime. As a comparison, {we also plot the curve of the fidelity $\mathcal{F}_R:=|\langle G|G_R\rangle|^2$, where $|G_R\rangle=|0_A,0_B,g\rangle$ is the ground state under the RWA}. It shows that $\mathcal{F}_R$ decreases dramatically as $g$ increases. Therefore, the TRWA gives a more accurate analytical result, at least for the ground state of the two-mode Rabi model.

\section{Detection of $g$ via the auxiliary resonator}
\label{detect}

\begin{figure}[tbp]
\centering
\includegraphics[width=1\columnwidth]{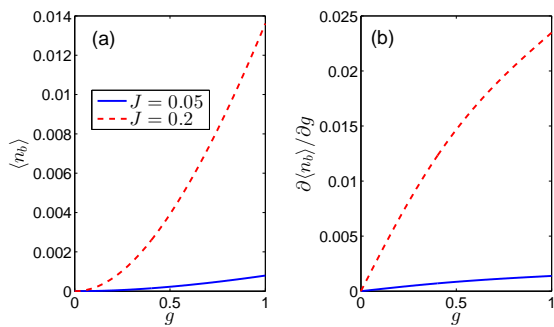}
\caption{(Color online) (a) The average photon number and (b) its derivation with respect to $g$.  The parameters are set as $\Omega=0.1$ and $\omega=1$.}
\label{metrology}
\end{figure}

{In the above section, we have shown that the system will acquire some excitations for the bosonic modes even in the ground state due to the {counter rotating wave coupling terms between the emitter and the resonator}. It is intuitive that the coupling strength can be detected by {directly} measuring the average excitation number in the resonator. However, a direct measurement on the resonator will {undoubtedly} disturb the resonator-emitter coupling system, and we choose to detect the coupling strength between the resonator $a$ and the emitter by measuring the average excitation number in the auxiliary resonator $b$. In other words, we regard the auxiliary resonator as the ``measurement apparatus'', and the photonic hopping induces the interaction between the system and {the measurement apparatus.}

{According to Eq.~(\ref{ground}), the average excitation number in the auxiliary resonator $b$ under the ground state is expressed as}
\begin{equation}
\langle n_b\rangle=\langle G_T|b^\dagger b|G_T\rangle=\frac{(\xi_A-\xi_B)^2}{2}.
\label{nb}
\end{equation}

In Fig.~\ref{metrology}(a), we plot $\langle n_b\rangle$ as a function of the coupling strength $g$. As shown in the figure, the average excitation number of the auxiliary resonator is a monotropic function of $g$, {implying that} the auxiliary resonator can be applied to perform the detection for the coupling strength $g$. The sensitivity of the detection {can be} evaluated by {the slope of  $\langle n_b\rangle$ with respect to $g$} and we regard it as a better detection when the slope is larger. According to Eq.~(\ref{nb}), the slope can be expressed as
\begin{equation}
\frac{\partial \langle n_b\rangle}{\partial g}=(\frac{\partial\xi_A}{\partial g}-\frac{\partial\xi_B}{\partial g})(\xi_A-\xi_B),
\end{equation}
where $\partial\xi_{A(B)}/\partial g$ can be determined from Eq.~(\ref{parameter}).
{In Fig.~\ref{metrology}(b), we plot the slope $\partial \langle n_b\rangle/\partial g$, as a function of the coupling strength $g$.} It is shown that a large slope can be achieved by both increasing $J$ and $g$, so as to be beneficial for performing a more sensitive quantum detection.

\section{Conclusion}
\label{conclusion}
In this paper, we have investigated the ground state of the two-mode quantum Rabi model, where {the} auxiliary resonator $b$ serves as a sensor for the coupling strength $g$ between the two-level emitter and the resonator $a$. In the large detuning or/and ultra-strong coupling regime, {we obtain an analytical result for the ground state based on the TRWA. The higher fidelity of our results comparing with the usual RWA is rooted in that the displacement induced by the ultra-strong coupling lowers the ground state energy. Moreover, we propose an indirect
detection scheme for the emitter-resonator coupling strength $g$ by
monitoring the average excitation number in the auxiliary resonator. We find that, the average excitation number is a monotropic function of $g$, and its slope with respect to $g$ will increase with the inter-resonator coupling strength}. We hope that our studies can be generalized to investigate the property of ground state for a more complicated photonic hybrid system in the ultra-strong coupling regime. Also, the discussions about the quantum detection by introducing the auxiliary resonator in this paper may be useful in the field of quantum sensing and quantum metrology.

\begin{acknowledgments}
 This work is supported by the Jilin province science and technology development plan item (under Grant No.~20170520132JH) and the NSFC (under Grant No.~11774024, No.~11534002, No.~U1530401).
\end{acknowledgments}

\end{document}